\documentclass[12pt]{article}
\usepackage{amsmath}
\usepackage{amssymb}
\usepackage{amsthm}
\usepackage{fullpage}
\usepackage[utf8]{inputenc}
\usepackage{mathtools}
\usepackage{url}
\usepackage[usenames,svgnames]{xcolor}

\usepackage[pagebackref=false]{hyperref} 
\usepackage[all]{hypcap} 

\DeclareMathOperator{\cyc}{cyc}

\DeclareMathOperator{\Li}{Li}

\theoremstyle{plain}

\newtheorem{theorem}{Theorem}[section]
\newtheorem{proposition}{Proposition}[section] 
\newtheorem{corollary}[theorem]{Corollary}

\newtheorem{remark}{Remark}[section]

\numberwithin{equation}{section} 

\hypersetup{
  breaklinks,colorlinks,
  citecolor=Green,
  linkcolor=BlueViolet,
  urlcolor=DarkBlue,
  pdftitle={Multispecies Quantum Hurwitz numbers },
  pdfauthor={J. Harnad and A. Yu. Orlov},
}

\def\be{\begin{equation}}
\def\ee{\end{equation}}

\def\bea{\begin{eqnarray}}
\def\eea{\end{eqnarray}}

\def\bt{\begin{theorem}}
\def\et{\end{theorem}}

\def\bp{\begin{proposition}}
\def\ep{\end{proposition}}

\def\bc{\begin{corollary}}
\def\ec{\end{corollary}}

\def\br{\begin{remark}\rm\small}
\def\er{\end{remark}}

\def\ss{\subset}
\def\deq{\coloneqq}

\def\ss{\subset}

\def\&{&{\hskip -20pt}}

\def\JJ{\mathcal{J}}

\def\Cb{\mathbf{C}}

\def\Nb{\mathbf{N}}

\def\Zb{\mathbf{Z}}

\begin{document}
\baselineskip 16pt
\medskip
\begin{center}
\begin{Large}\fontfamily{cmss}
\fontsize{17pt}{27pt}
\selectfont
\textbf{Multispecies quantum Hurwitz numbers}\footnote{Work  supported by the Natural Sciences and Engineering Research Council of Canada (NSERC) and the Fonds de recherche du Qu\'ebec - Nature et technologies (FRQNT)}
\end{Large}\\
\bigskip
\begin{large} {J. Harnad}$^{1,2}$ 
 \end{large}
\\
\bigskip
\begin{small}
$^{1}${ \em Centre de recherches math\'ematiques,
Universit\'e de Montr\'eal\\ C.~P.~6128, succ. centre ville,
Montr\'eal,
QC, Canada H3C 3J7} \\
 e-mail: harnad@crm.umontreal.ca \\
\smallskip
$^{2}${ \em Department of Mathematics and
Statistics, Concordia University\\ 7141 Sherbrooke W.,
Montr\'eal, QC
Canada H4B 1R6} 
\end{small}
\end{center}
\bigskip

\begin{abstract}
The construction of hypergeometric 2D Toda $\tau$-functions
as generating functions for quantum  Hurwitz numbers is extended here
 to multispecies families.  Both the enumerative geometrical significance  of these  
 multispecies weighted  Hurwitz numbers as  weighted enumerations of 
 branched coverings of the Riemann sphere and their combinatorial 
significance in terms of weighted paths in the Cayley graph of $S_n$ are derived.

\end{abstract}


\section{Introduction}

\subsection[weightedHurwitz]
{ Weighted Hurwitz numbers }

Quantum weighted Hurwitz numbers were introduced in \cite{GH2},  in four variants, and  a $1$-parameter family of
$2D$ Toda $\tau$-function  generating functions was constructed for each.
In the first case, the weight generating function  is:
\bea
E(q,z) &\&:= \prod_{i=0}^\infty (1+q^i z) = 1 +\sum_{i=1}^\infty E_i(q)z^i , \\
E_i(q)&\&:= \left(\prod_{j=1}^j {q^{j-1} \over 1 - q^j}\right).
\eea
The second is a slight modification of this, with weight generating function 
\bea
E'(q,z) &\&:= \prod_{i=1}^\infty (1+q^i z) = 1 +\sum_{i=1}^\infty E'_i(q)z^i , \\
E'_i(q)&\&:= \left(\prod_{j=1}^j {q^i \over 1 - q^j}\right).
\eea
The third case is based on the weight generating function
\bea
H(q,z) &\&:= \prod_{i=1}^\infty(1-q^i z)^{-1} =1 +\sum_{i=1}^\infty H_i(q)z^i , \\
H_i(q)&\&:= \left(\prod_{j=1}^i {1 \over 1 - q^j}\right),
\eea
and the final case is a hybrid, formed from the product  of the first and third 
for two distinct quantum deformation parameters
$q$ and $p$, with weight generating function
\bea
Q(q, p, z)  &\&:= \prod_{k=0}^\infty (1+ q^k z) (1- p^k z)^{-1} = \sum_{i=0}^\infty Q_i(q,p)z^i, \\
Q_i(q,p) &\& := \sum_{m=0}^i q^{\frac{1}{2}m(m-1)} \left(\prod_{j=1}^m (1-q^j) \prod_{j=1}^{i -m}(1-p^j)\right)^{-1},
\quad Q_\lambda(q,p) =\prod_{i=1}^{\ell(\lambda)} Q_{\lambda_i}(q,p), 
\eea
These are all expressible as exponentials  of the quantum dilogarithm function
\bea
\Li_2(q, z) &\&\deq \sum_{k=1}^\infty \frac{z^k}{k (1- q^k)}.
\label{quanumt_dilog} \\
 E(q, z) &\&= e^{-\Li_2(q, -z)}, \quad   E'(q, z) = (1+z)^{-1}e^{-\Li_2(q, -z)} \\
 H(q, z)  &\& = e^{\Li_2(q, z)}, \quad   Q(q, p, z)  = e^{\Li_2(p, z)-\Li_2(q, -z)}, \quad .
\eea

   The  content products for the first and third of these are defined to be
\bea
r_\lambda^{E(q,z)}(N)  &\& := \prod_{(ij) \in \lambda} E(q, (N+j-i)z) \\
r_\lambda^{H(q,z)}(N)  &\& := \prod_{(ij) \in \lambda} H(q, (N+ j-i)z).
\eea
The associated hypergeometric 2D Toda $\tau$-functions have
diagonal double Schur function expansions with these as coefficients:
\bea
\tau^{E(q,z)}(N, {\bf t}, {\bf s}) &\&=  \sum_{\lambda} r_\lambda^{E(q,z)} (N) S_\lambda({\bf t}) S_\lambda({\bf s}) \\
\tau^{H(q,z)}(N, {\bf t}, {\bf s}) &\&=  \sum_{\lambda} r_\lambda^{H(q,z)} (N) S_\lambda({\bf t}) S_\lambda({\bf s)}). 
\label{hypergeometric_quantum_expansion}
\eea

Using the Frobenius character formula \cite{Mac},  
\be
S_\lambda = \sum_{\mu, \ |\mu| = |\lambda|} {\chi_\lambda P_\mu \over Z_\mu}
\label{Frobenius_character}
\ee
and setting $N=0$, these may be rewritten as double expansions
 in the power sum symmetric functions  \cite{GH2}:
 \bea
\tau^{E(q,z)}({\bf t}, {\bf s})&\& := \tau^{E(q,z)}(0, {\bf t}, {\bf s})=\sum_{d=0}^\infty z^d\sum_{\mu, \nu, \, |\mu|=|\nu|}  H^d_{E(q)}(\mu, \nu)   P_\mu({\bf t}) P_\nu({\bf s}),
 \\
\tau^{H(q,z)}({\bf t}, {\bf s}) &\& :=\tau^{H(q,z)}(0, {\bf t}, {\bf s}) = \sum_{d=0}^\infty z^d \sum_{\mu, \nu, \, |\mu|=|\nu|}  H^d_{H(q)}(\mu, \nu)  P_\mu({\bf t}) P_\nu({\bf s}).
\label{quantum_hurwitz_expansion}
\eea
The coefficients are the corresponding quantum Hurwitz numbers $H^d_{E(q)}(\mu, \nu)$, $H^d_{H(q)}(\mu, \nu)$,
 which count weighted $n$-sheeted branched coverings of the Riemann sphere,   defined by
\bea
H^d_{E(q)}(\mu, \nu) := \sum_{k=0}^\infty  \sideset{}{'}\sum_{\mu^{(1)}, \dots \mu^{(k)} \atop \sum_{i=1}^k \ell^*(\mu^{(i)} )= d} 
 W_{E(q)}(\mu^{(1)}, \dots , \mu^{(k)})  H(\mu^{(1)}, \dots , \mu^{(k)}, \mu, \nu) ,
   \label{Hd_Eq} \\
H^d_{H(q)}(\mu, \nu) := \sum_{k=0}^\infty   (-1)^{k+d} \sideset{}{'}\sum_{\mu^{(1)}, \dots \mu^{(k)} \atop \sum_{i=1}^k \ell^*(\mu^{(i)} )= d} 
 W_{H(q)}(\mu^{(1)}, \dots , \mu^{(k)})  H(\mu^{(1)}, \dots , \mu^{(k)}, \mu, \nu).
  \label{Hd_Hq}
\eea
where the weightings for such covers with $k$ additional branch points are:
\bea
W_{E(q)} (\mu^{(1)}, \dots, \mu^{(k)}) 
&\&={1\over k!} \sum_{\sigma\in S_k}   {q^{(k-1) \ell^*(\mu^{(1)})} \cdots  q^{ \ell^*(\mu^{(k-1)})} \over 
 (1- q^{\ell^*(\mu^{(\sigma(1)})} )  \cdots (1- q^{\ell^*(\mu^{(\sigma(1)})} \cdots q^{\ell^*(\mu^{(\sigma(k)})})},
 \label{W_E_q}   \\
 W_{H(q)} (\mu^{(1)}, \dots, \mu^{(k)}) &\&= {1\over k!}\sum_{\sigma\in S_k}   {1 \over 
 (1- q^{\ell^*(\mu^{(\sigma(1)})} )  \cdots (1- q^{\ell^*(\mu^{(\sigma(1)})} \cdots q^{\ell^*(\mu^{(\sigma(k)})})}
 \label{W_H_q}   .
   \eea
Here
\be 
\ell^*(\mu) := |\mu| - \ell(\mu)
\ee
is the {\it colength} of the partition $\mu$, which is the index of coalescence of the sheets of the branched
cover over the branch points with  ramification profiles $\mu^{(i)}$,  the sum $\sum'_{\mu^{(1)}, \dots \mu^{(k)} \atop \sum_{i=1}^k \ell^*(\mu^{(i)} )= d} $ is over all $k$-tuples of partitions  having nontrivial ramification profiles that satisfy  the constraint $\sum_{i=1}^k \ell^*(\mu^{(i)} )= d$,  and $H(\mu^{(1)}, \dots , \mu^{(k)}, \mu, \nu)$ is the number of branched $n$-sheeted coverings, up to isomorphism,  having $k+2$ branch points with ramification profiles $(\mu^{(1)}, \dots , \mu^{(k)}, \mu, \nu)$.

These thus count weighted covers with a pair of branch points, say  $(0, \infty)$, having
ramification profiles of type $(\mu, \nu)$  and an  arbitrary number
of further  branch points, whose profiles $(\mu^{(1)}, \dots , \mu^{(k)})$ are constrained  only  by the requirement that the sum of the colengths,
which is related to the genus by the Riemann-Hurwitz formula
\be
\sum_{i=1}^k \ell^*(\mu^{(i)} )= 2g -2  +\ell(\mu) +\ell(\nu) =d,
\label{riemann_hurwitz}
\ee
be fixed to equal $d$.  

Another interpretation that is purely combinatorial can be given to the quantum Hurwitz numbers
 $ H^d_{E(q)}(\mu, \nu)$ and $ H^d_{H(q)}(\mu, \nu)$  appearing in ({\ref{quantum_hurwitz_expansion}), as
 follows. Let $(a_1b_1) \cdots (a_d b_d)$ be a product of $d$ transpositions $(a_i b_i) \in S_n$ in the symmetric
group  $S_n$ with $a_i < b_i$, $i=1, \dots, d$.
If $h \in  S_n$ is in the conjugacy class $ \cyc(\mu)\ss S_n$ consisting of elements with cycle lengths 
equal to the parts $(\mu_1, \dots , \mu_{\ell(\mu)})$ of the partition $\mu$ of weight $|\mu|=n$,
and length $\ell(\mu)$, we may view the successive steps in the product
\be
(a_1b_1) \cdots (a_d b_d) h
\ee
as a path in the Cayley graph generated by all transpositions. To each such path, we attach a {\it signature}
consisting of a partition $\lambda$ of $d$, $|\lambda|=d$, whose parts $\lambda_i$ consist
of the number of transpositions $(a_i b_i)$ sharing the same second element. If we further
require that the ones with equal second elements be grouped together into consecutive subsequences in which these second elements
are constant,  with the consecutive subsequences  strictly increasing in their second elements, then the number $\tilde{N}_\lambda$
of elements with signature $\lambda$ is related to the number $N_\lambda$ of such ordered sequences by
\be
\tilde{N}_\lambda = {|\lambda|! \over \prod_{i=1}^{\ell(\lambda)} \lambda_i !} N_\lambda
\ee
Denote the number of such paths from the conjugacy class of cycle type $\cyc(\mu)$ to the one of type $\cyc(\nu)$
having signature $\lambda$ as $\tilde{m}^\lambda_{\mu \nu}$, and the number of ordered sequences
of type $\lambda$ as  $m^\lambda_{\mu \nu}$. Thus
\be
\tilde{m}^\lambda_{\mu \nu} =  {|\lambda|! \over \prod_{i=1}^{\ell(\lambda)} \lambda_i !}  m^\lambda_{\mu \nu}.
\ee

For  all  paths of signature $\lambda$ we  assign weights
\bea
 \tilde{E}_\lambda(q) &\&=  \prod_{i=1}^{\ell(\lambda)}\lambda_i! E_{\lambda_i}(q)
 =\prod_{i=1}^{\ell(\lambda)}{\lambda_i! q^{{1\over 2}\lambda_i(\lambda_i -1)} \over \prod_{j=1}^{\lambda_i} (1-q^j)} , \\
 \tilde{H}_\lambda(q) &\&=  \prod_{i=1}^{\ell(\lambda)} \lambda_i! H_{\lambda_i}(q)
 =\prod_{i=1}^{\ell(\lambda)}{\lambda_i!\over \prod_{j=1}^{\lambda_i} (1-q^j)} 
\eea
to paths in the Cayley graph, and a pair of corresponding combinatorial weighted Hurwitz numbers
\bea
F^d_{E(q)} (\mu, \nu) &\&:= {1\over d!} \sum_{\lambda, \, |\lambda|=d} \tilde{E}_\lambda \tilde{m}^\lambda_{\mu \nu}, \\
F^d_{H(q)} (\mu, \nu) &\&:= {1\over d!} \sum_{\lambda\, |\lambda|=d} \tilde{H}_\lambda \tilde{m}^\lambda_{\mu \nu},
\eea
that give the weighted enumeration of paths, using the weighting factors $ \tilde{E}_\lambda(q)$ and $ \tilde{H}_\lambda(q) $
respectively for all paths of signature $\lambda$.

It was shown in \cite{GH2} that the enumerative geometrical and combinatorial definitions
of these quantum weighted Hurwitz numbers coincide:
\be
H^d_{E(q)}(\mu, \nu)  = F^d_{E(q)}(\mu, \nu),  \quad H^d_{H(q)}(\mu, \nu) = F^d_{H(q)}(\mu, \nu).
\label{Hd_equals_Fd}
\ee
A similar result holds for weights generated by the function $E'(q,z)$, with corresponding quantum Hurwitz numbers
defined by
\be
H^d_{E'(q)}(\mu, \nu) := \sum_{k=0}^\infty  \sideset{}{'}\sum_{\mu^{(1)}, \dots \mu^{(k)} \atop \sum_{i=1}^k \ell^*(\mu^{(i)} )= d} 
 W_{E'(q)}(\mu^{(1)}, \dots , \mu^{(k)})  H(\mu^{(1)}, \dots , \mu^{(k)}, \mu, \nu) ,
   \label{Hd_E'q} 
\ee
where the weights $W_{E'(q)} (\mu^{(1)}, \dots, \mu^{(k)})$ are given by
\bea
W_{E'(q)} (\mu^{(1)}, \dots, \mu^{(k)}) &\&:={1\over k!} \sum_{\sigma\in S_k}   {q^{(k) \ell^*(\mu^{(1)})} \cdots  q^{ \ell^*(\mu^{(k)})} \over 
 (1- q^{\ell^*(\mu^{(\sigma(1)})} )  \cdots (1- q^{\ell^*(\mu^{(\sigma(1)})} \cdots q^{\ell^*(\mu^{(\sigma(k)})})},\\
 &\&:={1\over k!} \sum_{\sigma\in S_k}   {1 \over 
 (q^{-\ell^*(\mu^{(\sigma(1)})}-1  )  \cdots (q^{-\ell^*(\mu^{(\sigma(1)})} \cdots q^{-\ell^*(\mu^{(\sigma(k)})}-1 )}.
 \label{W_E'_q} 
 \eea
Choosing $q$ as a positive real number, parametrizing it as
\be
q= e^{- \beta \hbar \omega},  \quad \beta={1\over kT}
\ee
and identifying the energy levels $\epsilon_k$   as those for a Bose gas with  linear spectrum in the
integers, as for a 1-D harmonic oscillator
\be
\epsilon_k := k \hbar \omega, \quad k\in \Nb,
\ee
we see that if we assign the energy
\be
\epsilon(\mu) := \epsilon_{\ell^*(\mu)} = \ell^*(\mu)\hbar \omega
\ee
to each branch point with ramification profile of type $\mu$, it contributes a factor
\be
n(\mu) ={1 \over e^{\beta\epsilon(\mu)} -1}
\ee
to the weighting distributions, the same as that for a bosonic gas.

\section[Generating functions for multispecies weighted Hurwitz numbers]
{2D Toda $\tau$-functions as generating functions  for \\ multispecies weighted Hurwitz numbers}

\subsection[The 2D Toda $\tau$-functions $\tau^{G({\bf q}, {\bf z}, {\bf z})}(N,{\bf t}, {\bf s})$]
{The multiparameter family  of $\tau$-functions $\tau^{G( {\bf z}, {\bf w})} (N,{\bf t}, {\bf s})$}

We now consider a multiparameter family of weight generating functions $G( {\bf z}; {\bf w}) $ obtained by
taking the product of any number of  generating functions $G_i( z_i)$ and $\tilde{G}_j (w_j)$
for distinct sets  of generating function parameters $ {\bf z} = (z_1, \dots , z_l)$, ${\bf w} = (w_1, \dots , w_m)$.
\be
Q({\bf q}, {\bf z}; {\bf p}, {\bf w}) := \prod_{i=1}^l E(q_i, z_i)  \prod_{j=1}^m H(p_j, w_j).
\ee

   Following the approach developed in \cite{GH2}, we define an associated element of the
   center $\Zb(\Cb[S_n])$ of the group algebra $\Cb[S_n]$ by
   \be
  \hat{Q}({\bf q}, {\bf z}; {\bf p}, {\bf w}, \JJ) := \prod_{a=1}^n Q({\bf q},  \JJ_a {\bf z}; {\bf p},  \JJ_a {\bf w}) , 
   \ee
where $\JJ:= (\JJ_1, \dots, \JJ_n)$ are the Jucys-Murphy elements \cite{Ju, Mu, DG}
\be
\JJ_1:= 0,  \ \JJ_b :=\sum_{a=1}^{b-1} (a b), \quad b=1, \dots , n,
\ee
which generate an abelian subalgebra of $\Zb(\Cb[S_n])$. The element $\hat{Q}({\bf q}, {\bf z}; {\bf p}, {\bf w}, \JJ)$  defines 
an endomorphism of $\Zb(\Cb[S_n])$ under multiplication, which is diagonal in the basis  $\{F_\lambda\}$
of $\Zb(\Cb[S_n])$ consisting of the orthogonal idempotents, corresponding to irreducible representations, labelled by partitions  
$\lambda$ of $n$:
\be
\hat{Q}({\bf q}, {\bf z}; {\bf p}, {\bf w}, \JJ) F_\lambda = r_\lambda^{Q({\bf q},  {\bf z}; {\bf p},  {\bf w})} F_\lambda
\label{central_qpzw_generator}
 \ee
 where
 \be
r_\lambda^{Q({\bf q},  {\bf z}; {\bf p},  {\bf w})} =  \prod _{i=1}^l r_\lambda^{E(q_i)} (z_i)    \prod_{j=1}^m r_\lambda^{H(p_j)} (w_j).
\ee

More generally, defining 
 \be
r_\lambda^{Q({\bf q},  {\bf z}; {\bf p},  {\bf w})} (N) =  \prod _{i=1}^l r_\lambda^{E(q_i)} (N, z_i)    \prod_{j=1}^m
 r_\lambda^{H(p_j)} (N, w_j),
\ee
where
\bea
r_\lambda^{E(q)} (N, z) &\& := \prod_{(ij) \in \lambda} E(q, (N+ j-i)z) \\
r_\lambda^{H(q)}(N, z)  &\& := \prod_{(ij) \in \lambda} H(q, (N+ j-i)z), 
\eea
we have
\be
r_\lambda^{Q({\bf q},  {\bf z}; {\bf p},  {\bf w})}  =  r_\lambda^{Q({\bf q},  {\bf z}; {\bf p},  {\bf w})} (0) .
\ee
From general considerations \cite{OrSc, HOr},   the double Schur function series
\be
\tau^{Q({\bf q}, {\bf z}; {\bf p},  {\bf w})}(N, {\bf t}, {\bf s}) := \sum_\lambda  r_\lambda^{Q({\bf q},  {\bf z}; {\bf p},  {\bf w})} (N) 
\, S_\lambda({\bf t}) S_\lambda({\bf s})
\label{tau_qz_pw}
\ee
is known to define a family of $2D$ Toda $\tau$-functions of hypergeometric type.

\subsection[Multiparametric_families_geometric]
{ Multispecies geometric  weighted  Hurwitz numbers }

We now consider coverings in which the branch points are divided into two different classes,
$\{\mu_+^{i, u_i)}\}_{i=1, \dots, l; \,  u_i = 1 \dots , k^-_i}$ and $\{\mu_-^{j, v_j)}\}_{j=1, \dots, m; \, v_j = 1, \dots , k^-_j}$ 
corresponding to the weight generating functions of type $E(q_i)$ and $H(p_j)$ respectively, each of which is
further divided into $l$ and $m$ distinct species (or ``colours''), of which there are $\{k^+_i\}$ and 
$\{k^-_j\}$ branch points of types $E$ and $H$ and colours $i$ and $j$ respectively. The weighted number of such coverings 
 with specified total colengths    ${\bf c} =(c_1, \dots , c_l),  \quad {\bf d} =(d_1, \dots , d_m),  \quad c_i, d_j \in \Nb$ for each 
class and colour is  the multispeciesl quantum Hurwitz number
\bea
&\&\phantom{.} {\hskip 60 pt}H^{({\bf c}, {\bf d})}_{({\bf q}, {\bf p})}(\mu, \nu)  := 
  \sum_{\{k_i^+=1; k_j^-=1\} \atop i=1, \dots, l; j=1. \dots, m}^\infty 
 \sum_{\{\mu_+^{( i, u_i)}\});\  \mu_-^{(j, v_j)}\}\atop \sum_{u_i=1}^{k^+_i}\ell^*(\mu_+^{(i, u_i)} )= c_i,
  \ \sum_{v_j=1}^{k^-_j}\ell^*(\mu_-^{(j, v_j)} )= d_j }
  \cr
&\&
\times  \prod_{i=1}^l W_{E(q_i)}(\mu_+^{(i,1)}, \dots , \mu_+^{(i, k^+_i)})  \prod_{j=1}^m W_{H(p_j)}(\mu_-^{(j,1)}, \dots , \mu_-^{(j, k^-_j)})
  H(\{\mu_+^{(i, u_i) };\  \mu_-^{(j,  v_ j)}\}, \mu, \nu ). \cr
  &\&
     \eea
    
Substituting the Frobenius-Schur formula (\ref{Frobenius_Schur_Hurwitz}) and the
Frobenius character formula (\ref{Frobenius_character}) 
into (\ref{tau_qz_pw}), it follows that  $\tau_{Q({\bf q}, {\bf z}; {\bf p},  {\bf w})}(N, {\bf t}, {\bf s})$
is the generating function for  $H^{({\bf c}, {\bf d})}_{({\bf q}, {\bf p})}(\mu, \nu) $.
Using multi-index notion to denote
\be
\prod_{i=1}^l z_i^{c_i} \prod_{j=1}^m w_j^{d_j} =:  {\bf z}^{\bf c} {\bf w}^{\bf d}, \quad 
  \ee
  we have:
\bt
\be
\tau^{Q({\bf q}, {\bf z}; {\bf p},  {\bf w}}(0, {\bf t}, {\bf s})  :=\sum_{{\bf c}=(0, \dots,0);\ {\bf d} = (0, \dots, 0)}^ {(\infty, \dots, \infty)}
{\hskip -20 pt} {\bf z}^{\bf c} {\bf w}^{\bf d}\sum_{\mu, \nu}
 H^{({\bf c}, {\bf d})}_{({\bf q}, {\bf p})}(\mu, \nu)P_\mu({\bf t}) P_\nu({\bf s}). 
 \label{multispecies_geometric_tau}
 \ee
 \et

\subsection[Multiparametric_families_combinatorial]
{ Multispecies  combinatorial weighted Hurwitz numbers }

Another basis for $\Zb(\Cb[S_n])$  consists of the cycle sums
\be
C_\mu := \sum_{h\in cyc(\mu)} h,
\ee
 where $cyc(\mu)$ denotes the conjugacy class  of elements $h\in \cyc(\mu)$ with cycle lengths equal
to the parts of $\mu$. 
 The two  are related by
\be
F_\lambda = h_\lambda^{-1} \sum_{\mu, \ |\mu| = |\lambda|} \chi_\lambda(\mu) C_\mu, 
\label{F_lambda_C_mu}
\ee
where $\chi_{\lambda}(\mu)$ denotes the irreducible character of the irreducible
representation of type  $\lambda$ evaluated on the conjugacy class $\cyc(\mu)$. 
Under the characteristic map, this   is equivalent to the Frobenius character formula (\ref{Frobenius_character}).
The Macmahon generating function 
\be
\prod_{i=1}^\infty (1-q^i)^{-1} = \sum_{n=0}^\infty D_n q^n,
\ee
gives the number  $D_n$  of partiitions of $n$. We denote by ${\bf F}^{(n, c_i)}_{E(q_i)}$ and ${\bf F}^{(n, d_j)}_{H(q_j)}$ 
 the $D_n \times D_n$ matrices, whose  elements are $\left(F^{c_i}_{E(q_i)}(\mu, \nu)\right)_{|\mu|=|\nu|= n}$
 and  $\left(F^{d_j}_{E(q_j)}(\mu, \nu)\right)_{|\mu|=|\nu|= n}$, respectively, for $i=1, \dots, l$, $j=1, \dots, m$.
Since these  represent central elements of the group algebra $\Cb[S_n]$, they commute amongst themselves.
Denoting the matrix product:
\be
{\bf F}^{({\bf c}, {\bf d})}_{({\bf q}, {\bf p})} = \prod_{i=1}^l {\bf F}^{(n, c_i)}_{E(q_i)}   \prod_{j=1}^m {\bf F}^{(n, d_j)}_{Hq_j)} ,
\ee
its matrix elements, denoted  $F^{({\bf c}, {\bf d})}_{({\bf q}, {\bf p})}  (\mu, \nu)$,  may be interpreted as the weighted
number of
\be
d := \sum_{i=1}^lc_i + \sum_{j=1}^m d_j 
\ee
step paths in the Cayley graph from the conjugacy class of cycle  type $\mu$ to the one of type  $\nu$,
where all paths are divided into equivalence classes, according to their multisignature.
This consists of  partitions of  $\lambda_+^{(i)}$ of $c_i$ and $\lambda_-^{(j)}$   of $d_j$ with
\be
\lambda_+^{(i)}= \ell( \lambda_+^{(i)}), \quad \lambda_+^{(i)}= \ell( \lambda_-^{(j)})
\ee
parts  $\lambda_+^{(i)}$  and  $\lambda_+^{(i)}$', 
each of which is itself subdivided into parts $(\lambda_{+, u_i}^{(i)}, \lambda_{-,v_j, }^{(j)})$, in which the second elements
of the transpositions are constant, and distinct for each $\lambda_{+, u_i}^{(i)},$ or $\lambda_{-, v_j}^{(j)}$. 
We define the hypersignature of such a path as the set of numbers  
$\{\lambda_{+, u_i}^{(i)}, \lambda_{-, v_j}^{(j)}\}_{u_i =1, \dots, k^+_i. \ v_j = 1, \dots , k^-_j}$.
 The weight for each path within such an equivalence class is then the product of the weights for each subsegment:
\be
\prod_{i=1}^l \prod_{u_i=1}^{k^+_i}E(q_i)_{\lambda_{(+, u_i)}^{(i)}} \ \prod_{j=1}^m  \prod_{v_j=1}^{k^-_j} H(p_j)_{\lambda_{(-, v_j)}^{(j)}}
\ee
and $F^{({\bf c}, {\bf d})}_{({\bf q}, {\bf p}})  (\mu, \nu)$  is the sum of these, each  multiplied by the number
of elements of the equivalence class of paths. with the given hypersignature.  The multispecies generalization of (\ref{Hd_equals_Fd}) is 
equality of the geometric and combinatorial Hurwitz numbers:
\bt
\be
F^{({\bf c}, {\bf d})}_{({\bf q}, {\bf p})}  (\mu, \nu) = H^{({\bf c}, {\bf d})}_{({\bf q}, {\bf p})}  (\mu, \nu).
\ee
\et
\begin{proof}

Applying the central element (\ref{central_qpzw_generator}) 
to the cycle sum $C_\mu$ gives
\be
\hat{Q}({\bf q}, {\bf z}; {\bf p}, {\bf w}, \JJ)  C_\mu = \sum_{\nu, |\nu|=|\mu|}F^{({\bf c}, {\bf d})}_{({\bf q}, {\bf p}})  (\mu, \nu) \, C_\nu
\ee            
and the orthogonality of group characters that
\be
\tau^{Q({\bf q}, {\bf z}; {\bf p},  {\bf w})}({\bf t}, {\bf s})  :=\sum_{{\bf c}=(0, \dots,0);\ {\bf d} = (0, \dots, 0)}^ {(\infty, \dots, \infty)}
{\hskip -20 pt} {\bf z}^{\bf c} {\bf w}^{\bf d}\sum_{\mu, \nu}
 F^{({\bf c}, {\bf d})}_{({\bf q}, {\bf p})}(\mu, \nu)P_\mu({\bf t}) P_\nu({\bf s}). 
  \label{multispecies_combinatorial_tau}
 \ee
 Comparing this with eq.~(\ref{multispecies_geometric_tau}) proves the result.

\end{proof}

\br {\bf Multispecies Bose gases }

Returning to the interpretation of the quantum Hurwitz weighting distributions in terms
of Bose gases, if we choose each $q_i$ to be a positive real number with $q_i<1$, and
parametrize it, as before,
\be
q_i = e^{-\beta \hbar\omega_i}
\ee
for some ground state energy $\hbar \omega_i$, and again choose a linear energy spectrum,
with energy assigned to the branchpoint $\mu$ of type $i$ with profile type $\mu$ to be
\be
\epsilon^{(i)}(\mu) :=  \ell^*(\mu) \hbar \omega_i
\ee
 we see that the resulting contributions to the weighting distributions distributions
 from each species of branch points of ramification type $\mu^{(j)}$ are given by
 \be
n^{(i)}_{B}(\mu)=  {1\over e^{\beta \epsilon^{(i)}(\mu)}  -1},
 \label{multi_bose_distribution}
 \ee
 which are those of a multispecies  mixture of Bose gases.
 \er

\subsection[Examples]
{Examples of general multispecies Hurwitz numbers}

 \bigskip
\noindent {\it Acknowledgements.}  The author would  like to thank M. Guay-Paquet for helpful discussions.

\bigskip 

\newcommand{\arxiv}[1]{\href{http://arxiv.org/abs/#1}{arXiv:{#1}}}

\bigskip
\noindent

\end{document}